\documentclass[aps,pra,twocolumn,superscriptaddress,groupedaddress,longbibliography]{revtex4-1}
\usepackage{epsfig,amsopn}
\usepackage{graphicx}
\usepackage{color}
\usepackage{bbm}
\usepackage{amsmath,amssymb}
\usepackage{enumerate}
\usepackage{array}
\newcommand\bea{\begin{eqnarray}}
\newcommand\eea{\end{eqnarray}}
\newcommand\beq{\begin{equation}}
\newcommand\eeq{\end{equation}}

\def\nn{\nonumber}
\def\f{\frac}
\def\al{\alpha}

\def\de{\delta}
\def\ep{\epsilon}

\def\ga{\gamma}
\def\Ga{\Gamma}

\def\si{\sigma}

\def\De{\Delta}

\def\la{\langle}
\def\ra{\rangle}

\def\mPT{\mathcal{PT}}
\def\mP{\mathcal{P}}
\def\mT{\mathcal{T}}

\begin{document}
\title{Transmission across non-Hermitian $\mathcal{P}\mathcal{T}$-symmetric  quantum dots and ladders} 
 \author{ Abhiram Soori }  
 \email{abhirams@uohyd.ac.in}
 \affiliation{ School of Physics, University of Hyderabad, C. R. Rao Road, 
 Gachibowli, Hyderabad-500046, India.}
\author{ M. Sivakumar}
 \affiliation{ School of Physics, University of Hyderabad, C. R. Rao Road,
 Gachibowli, Hyderabad-500046, India.}
 \author{ V. Subrahmanyam}
 \affiliation{ School of Physics, University of Hyderabad, C. R. Rao Road,
 Gachibowli, Hyderabad-500046, India.}
 \begin{abstract}
A non-Hermitian region connected to semi-infinite Hermitian lattices acts either as a source or as a sink and the probability current is not conserved in a scattering typically. Even a $\mathcal{P}\mathcal{T}$-symmetric region that contains both a source and a sink does not lead to current conservation plainly.  We propose a model and study the scattering across  a non-Hermitian $\mathcal{P}\mathcal{T}$-symmetric two-level quantum dot~(QD) connected to  two semi-infinite one-dimensional lattices in a special way so  that the probability current is conserved. Aharonov-Bohm type phases are included in the model, which arise from magnetic fluxes ($\hbar\phi_{L} /e,~\hbar\phi_{R} /e$)  through two loops in the system.  We show that when $\phi_L=\phi_R$, the probability current is conserved. 
 We find that the transmission across the QD can be perfect in the $\mathcal{P}\mathcal{T}$-unbroken phase (corresponding to real eigenenergies of the isolated QD) whereas the transmission is never perfect in the $\mathcal{P}\mathcal{T}$-broken phase (corresponding to purely imaginary eigenenergies of the QD). The two transmission peaks have the same width only for special values of the fluxes  (being  odd multiples of $\pi\hbar/2e$). In the broken phase, the transmission peak is surprisingly not at zero energy. We give an insight into this feature through a four-site toy model. We extend the model to a $\mathcal{P}\mathcal{T}$-symmetric ladder connected to two semi-infinite lattices. We show that the transmission is perfect in unbroken phase of the ladder due to  Fabry-P\'erot type interference, that can be controlled by tuning the chemical potential. In the broken phase of the ladder, the transmission is substantially suppressed.
 \end{abstract}

\maketitle

\section{Introduction}
Conventional wisdom advocates the idea that real energy eigenvalues  in quantum mechanics  necessarily require   the Hamiltonian to be Hermitian. In 1998, Bender and Boettcher made a startling discovery that a non-Hermitian (NH) Hamiltonian can have real spectrum if it is $\mPT$-symmetric where $\mPT$ is the product of parity and time reversal operators~\cite{bender98,bender02,bender04}. 
When all the eigenenergies are real, the phase is termed $\mPT$-unbroken and when the eigenenergies are complex, the phase is termed $\mPT$-broken. Despite the fact that the parent Hamiltonian is invariant under $\mPT$, the eigenstates are not invariant under $\mPT$ in the $\mPT$-broken phase,  whereas in the $\mPT$-unbroken phase the eigenstates are invariant under $\mPT$.  The two phases are separated by a point in the parameter space called exceptional point, where not only the  eigenvalues are degenerate, even the eigen vectors coalesce. Moreover, non-Hermitian systems can also exhibit more than one exceptional point. The exceptional points can form a line, a ring~\cite{xu17} or a surface~\cite{zhang19}.  The  work by Bender et al. spawned a flurry of activities both in  theoretical~\cite{mosta02,mosta03,mosta10} and  experimental fronts~\cite{Ganai2018}. The latter span over a wide range of topics such as optics~\cite{guo09,ruter,kottos10,longhi17}, microwaves~\cite{bittner,sun14,poli15}, electronics~\cite{schindler}, acoustics~\cite{zhu14,au17}, optoelectronics~\cite{gao15}, superconducting transmon circuits~\cite{naghiloo19,chen21}, dissipative systems~\cite{nowak97} and localization transition~\cite{Hatano96}. Non-Hermitian skin effect wherein eigenstates pile up near the boundary of a non-Hermitian lattice has captured the attention of researchers in the community~\cite{yao18,song19,hoffman20,Liang22,franca22}.  Open systems with gain and loss can be modeled by NH Hamiltonians~\cite{rotter09}. In superconducting transmon systems, a non-Hermitian Hamiltonian can be realized by postselection~\cite{naghiloo19}. 
Realization of NH system in optics uses the mathematical equivalence between the one particle  Schr\"odinger equation  and the scalar electromagnetic  wave equation in the paraxial approximation,  with space dependent refractive index playing the role of the potential, and non-Hermiticity can be implemented through an absorptive part.
Investigations in this domain have led to interesting phenomena like unidirectional transmission~\cite{feng11,peng14,peng16}, improved sensors~\cite{wie16}, non-Hermitian topological sensors~\cite{budich20,chen17,rmp21} and single-mode lasers~\cite{feng14,hodaei14}. Topological invariants~\cite{ghatak2019}, chaotic dynamics~\cite{huang2020}  and entanglement~\cite{guo2021} have been studied in non-Hermitian systems. 

In a tight binding model, an imaginary onsite potential $i\ga$ acts as a source when $\ga>0$ and as a sink when  $\ga<0$~~\cite{Foa_Torres_2019,jin10}. A non-Hermitian  onsite potential can be viewed as an open system wherein the probability current can leak out (sink) or rush in (source)~\cite{jin10}.   Dynamics of a particle hopping on a non-Hermitian one dimensional lattice with imaginary onsite potentials corresponding to sink has been studied~\cite{Rudner09}. There has been some work on scattering across a non-Hermitian region in lattice~\cite{zhu16,shobe21} and continuum  models~\cite{levai01,levai02,deb03,cannata}. However, the current is not conserved in scattering across NH region in these studies.
Scattering in an infinite one-dimensional  $\mPT$-symmetric lattice across a central non-Hermitian region with two sites having onsite potentials $\pm i\ga$ and  coupled is studied, wherein the site with onsite potential  $i\ga$ ($-i\ga$) is connected only to the semi-infinite lattice on the left (right)~\cite{zhu16}. In this work, the transmission probabilities for particles incident from the left and from the right are equal since the particle experiences both the source and the sink irrespective of whether the particle is incident from left or right. However, a reflected particle incident from the left (right) experiences the source (sink) preferentially,  making the reflection probabilities different for particles incident from the left and the right. The probability current is not conserved in this model despite having both the source and the sink. But the scattering amplitudes satisfy a relation known as generalized unitarity in such systems~\cite{ge12,mosta14}. Further, unidirectional invisibility wherein reflection probability depends on the direction of propagation is possible in such systems~\cite{Ganai2018}. In some $\mPT$-symmetric systems, for a certain parameter regime, the current is known to be conserved~\cite{ahmed13}. Also, in certain $\mPT$-antisymmetric lattices, the current is found to be conserved~\cite{jin18}. $\mPT$-symmetric lattice models where the probability current is conserved have been studied~\cite{jin12,zhu15} in which the sites with onsite potentials $\pm i\ga$ are coupled to the rest of the lattice with equal weight. Probability current conservation is a special property which is usually not expected in scattering across a non-Hermitian region. We present a distinct, new related model in which the probability current is conserved.  

We propose a model where a non-Hermitian $\mPT$-symmetric quantum dot~(QD) is connected to two one-dimensional Hermitian lattices in such a way that the current is conserved in scattering across the $\mPT$-symmetric region. The $\mPT$-symmetric QD has two eigenenergies. When a QD in $\mPT$-unbroken phase is connected to the lattices, we find that the transmission probability can reach unity at two energies. When the QD at exceptional point is connected to the lattices, the transmission probability is unity for one value of energy.  On the other hand, when the QD in  $\mPT$-broken phase is connected to the lattices, the transmission probability is strictly less than unity at all energies. However, the transmission probability exhibits peak at two distinct (real) values of energy, despite the eigenenergies of the isolated QD being purely imaginary. We come up with a four-site toy model to explain this result. We also study the existence of the bound states in the system. The hopping amplitudes that connect the QD to the semi-infinite lattices can have a phase factor in general. This phase factor brings about an interference that is analogous to Aharonov-Bohm effect. We further extend the $\mPT$-symmetric QD to a $\mPT$-symmetric ladder, connect it to two semi-infinite lattices and study transmission. We find that the transmission probability exhibits Fabry-P\'erot type interference by tuning the chemical potential in the ladder region. Transmission probability exhibits distinct features for the $\mPT$-unbroken and $\mPT$-broken phases of the ladder.

The paper is structured as follows. In Sec.~\ref{sec-model}, we motivate and present the model proposed. This is followed by details of calculation and results with analysis.  In sec.~\ref{sec:ladder}, we extend the model of $\mPT$-symmetric quantum dot to $\mPT$-symmetric ladder and study scattering across the ladder. Finally, we discuss the implications of our results and conclude in sec.~\ref{sec-con}.
 \section{Scattering across $\mPT$-symmetric quantum dot}~\label{sec-model}
\subsection{Model and calculation}
 A minimal $\mPT$-symmetric model is defined by the Hamiltonian $H_c=i\ga\si_z-t'\si_x$, where $\si_x$, $\si_z$ are Pauli spin matrices, and the parameters $t',~\ga$ are real and positive. This can be thought of as a QD with two sites having onsite energies $\pm i\ga$ and a hopping $-t'$ between the sites. The eigenenergies of this model are \bea \ep_{\pm}=\pm\sqrt{t'^2-\ga^2}.\eea 
 The operator $\mP$ is $\si_x$ whereas the operator $\mT$ is complex conjugation making $\mPT H_c \mPT= H_c$. In the  regime $t'>\ga$, the eigenenergies are real, and the phase is termed $\mPT$-unbroken while for $\ga>t'$, the eigenenergies are imaginary and the phase is termed $\mPT$-broken. Scattering has been studied in a system where two semi-infinite one dimensional lattices are connected to the minimal $\mPT$-symmetric model~\cite{zhu16}. But,  in their work, the site with onsite energy $i\ga$  is connected to one lattice (labeled $L$) while the site with onsite energy $-i\ga$ is connected to the other lattice (labeled $R$). A particle incident on the QD  from the lattice $L$ experiences the source (the site with onsite energy $i\ga$) first  and then hops on to the sink (the site with onsite energy $-i\ga$) before hopping on to the lattice $R$. So, the effects of source and sink do not cancel out, and the incident current is not equal to the sum of currents carried by the particle reflected on the lattice $L$ and the particle transmitted onto the lattice $R$. We propose to connect the semi-infinite lattices to the $\mPT$-symmetric QD in such a way that the hopping strengths from the lattice onto both the sites of the QD have equal weight. In such a case, the particle incident from the lattice on to the QD is not favored towards either the source or sink and the current must be conserved. 
 
 \begin{figure}[htb]
  \includegraphics[width=8.81cm]{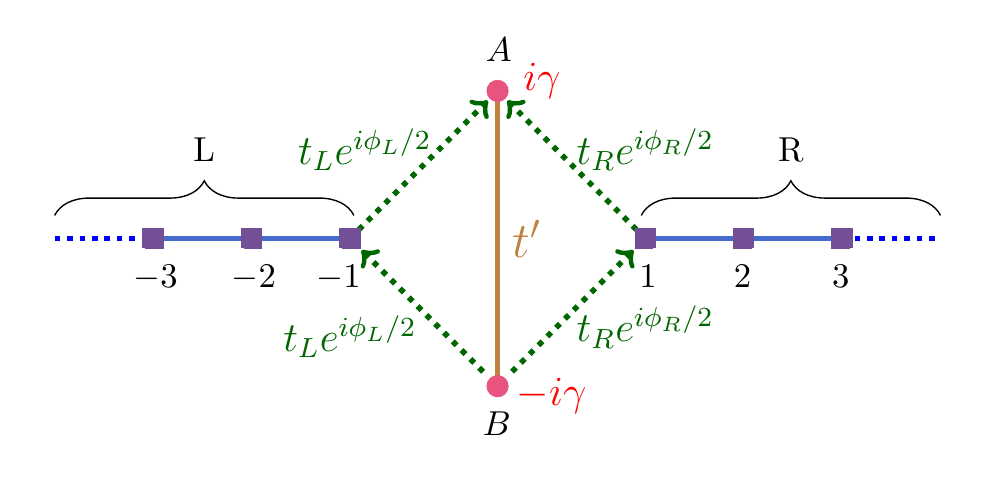}
  \caption{Schematic of semi-infinite lattices connected to non-Hermitian $\mPT$-symmetric quantum dot. }~\label{fig:schem}
 \end{figure}

 In the Hamiltonian $H_c$ for QD, the non-Hermitian part is proportional to $\si_z$. So, any additional Hermitian term that is a linear combination of $\si_x$, $\si_y$ and $\si_0$ will not break the $\mPT$ symmetry of $H_c$. The full Hamiltonian for the system is given by:
 \bea
 H &=& H_L+H_{QD}+H_R+H_{LQ}+H_{QR}, \nn \\
  && {\rm where}~~~H_L = -t\sum_{n=-\infty}^{-1}(c^{\dag}_{n-1}c_n+h.c.), \nn \\ 
  && H_{QD}=-t'(c^{\dag}_{A}c_B+h.c.)+i\ga(c^{\dag}_{A}c_A -c^{\dag}_{B}c_B), \nn \\
  && H_R = -t\sum_{n=1}^{\infty}(c^{\dag}_{n+1}c_n+h.c.), \nn \\
  && H_{LQ} = -t_L[c_{-1}^{\dag}(e^{-i\phi_L/2}c_A+e^{i\phi_L/2}c_B) + h.c.], \nn \\ 
  && H_{QR} = -t_R[c_1^{\dag}(e^{-i\phi_R/2}c_A+e^{i\phi_R/2}c_B)+h.c.].~~ \label{eq:ham}
 \eea
 Here, $t$ ($>t',~\ga$) is the hopping strength in the semi-infinite one dimensional lattices $L$ and $R$, $c_n$ and $c^{\dag}_n$ are annihilation and creation operators respectively at site labelled by $n$,  $t_{L/R}$ are real valued parameters that quantify the hopping from the two sites of the QD onto $L/R$, $\phi_{L/R}$ are the real-valued phase factors that accompany the hopping amplitudes as shown in eq.~\eqref{eq:ham}. Here, the lattice $L$ ($R$) is coupled to the linear combination $[e^{i\phi_L/2}|A\ra+e^{-i\phi_L/2}|B\ra]$ ($[e^{i\phi_R/2}|A\ra+e^{-i\phi_R/2}|B\ra]$) of the states $|A\ra$ and $|B\ra$ that constitute the QD. A schematic diagram of the setup under investigation is shown in Fig.~\ref{fig:schem}. The entire system is $\mPT$-symmetric only when $\phi_L=\phi_R$ and $t_L=r_R$ as will be discussed in sec~\ref{sec-con}.

 The dispersion relation in the semi-infinite lattices $L,~R$ is $E=-2t\cos{k}$. Scattering eigenstate of a  particle incident on the QD with energy $E$ from $L$ has   the form $|\psi\ra=\sum_{n}\psi_n|n\ra$ where  
 \bea
 \psi_{n} &=& e^{ikn}+\mathbbm{r}_E ~e^{-ikn},~~~{\rm for~~}n\le -1,\nn \\ 
 &=& \mathbbm{t}_E e^{ikn},~~~{\rm for~~}n\ge 1, \label{eq:psi} 
 \eea
and  $k=\cos^{-1}{(-E/2t)}$. The scattering coefficients $\mathbbm{r}_E$ and $\mathbbm{t}_E$ can be determined by using time independent Schr\"odinger equation.  $|\mathbbm{r}_E|^2$ and $|\mathbbm{t}_E|^2$ are the reflection and transmission probabilities respectively. Probability current at the bond $(n,n+1)$ is given by $J_{n,n+1}=-i\la\psi |(c^{\dag}_nc_{n+1}-c^{\dag}_{n+1}c_n) | \psi\ra$ for $n\ge1$ and $n\le-2$. The probability current is conserved when  $J_{-2,-1}=J_{1,2}$ for a steady state. From the form of the wave function in eq.~\eqref{eq:psi}, it can be seen that the probability current is conserved when $|\mathbbm{r}_E|^2+|\mathbbm{t}_E|^2=1$. We find that the probability current is conserved at all energies when $\phi_L=\phi_R$. When $t_L=t_R=t_d$, $\phi_L=\phi_R=\phi$, the expression for the scattering coefficients take the form: 
\bea  \mathbbm{r}_E= -\Big(\f{\Ga \cos{k}-t}{\Ga e^{ik}-t}\Big),~~~~~~ && \mathbbm{t}_E = \f{i~\Ga~ \sin{k}}{\Ga e^{ik}-t}, \label{eq:RT} \eea
where $\Ga=4t_d^2~(t'\cos{\phi}-E)/(E^2+\ga^2-t'^2)$. We set $t_L=t_R=t_d$ and $\phi_L=\phi_R=\phi$ except otherwise specified. 

The Hamiltonian in eq.~\eqref{eq:ham} combined with the Schr\"odinger wave equation implies that the eigenstate $|\psi\ra=\sum_n\psi_n|n\ra$ (where $n$ takes $A$, $B$ and nonzero integer values) obeys the equations: 
\bea 
E\psi_{-1} &=& -t\psi_{-2}-t_d(e^{-i\phi/2}\psi_A+e^{i\phi/2}\psi_B) \nn \\ 
E\psi_A&=&~i\ga\psi_A-t'\psi_B-t_de^{i\phi/2}(\psi_{-1}+\psi_1) \nn \\
E\psi_B&=&-i\ga\psi_B-t'\psi_A-t_de^{-i\phi/2}(\psi_{-1}+\psi_1) \nn \\
E\psi_{1} &=& -t\psi_{2}-t_d(e^{-i\phi/2}\psi_A+e^{i\phi/2}\psi_B) \label{eq:eom}
\eea
Substituting the form of scattering wavefunction $\psi_n$ given by eq.~\eqref{eq:psi} into eq.~\eqref{eq:eom} will result in four linear equations with four unknowns: $\mathbbm{r}_E$, $\mathbbm{t}_E$, $\psi_A$ and $\psi_B$. Solving these equations analytically gives the expressions for scattering amplitudes presented in eq.~\eqref{eq:RT}.

The phrase `quantum dot' is typically used to describe a small semiconductor heterostructure into and out of which electrons can hop. In this manuscript, QD is used to mean two sites $A,~B$ in which quantum particles that conform to the Hamiltonian $H_{QD}$ in eq.~\eqref{eq:ham} reside. The results presented in this work apply to both fermions and bosons that obey the Hamiltonians in eq.~\eqref{eq:ham} or eq.~\eqref{eq:ham2}. These results also apply to a system where metallic leads are connected to a quantum dot in which electrons are responsible for conduction. In such a system, a voltage bias $V$ applied across the quantum dot results in a current through the quantum dot. The transmission probability at energy $E$ is related to the differential conductance at voltage bias $V=E/e$ (where $e$ is the electron charge) by  Landauer-B\"uttiker formula~\cite{landauer1957r,buttiker1985m,datta1995} $G(V)=|\mathbbm{t}_E|^2(e^2/h)$.
 
\subsection{Current Conservation}~\label{sec-res}

\begin{figure}[htb]
 \includegraphics[width=8.0cm]{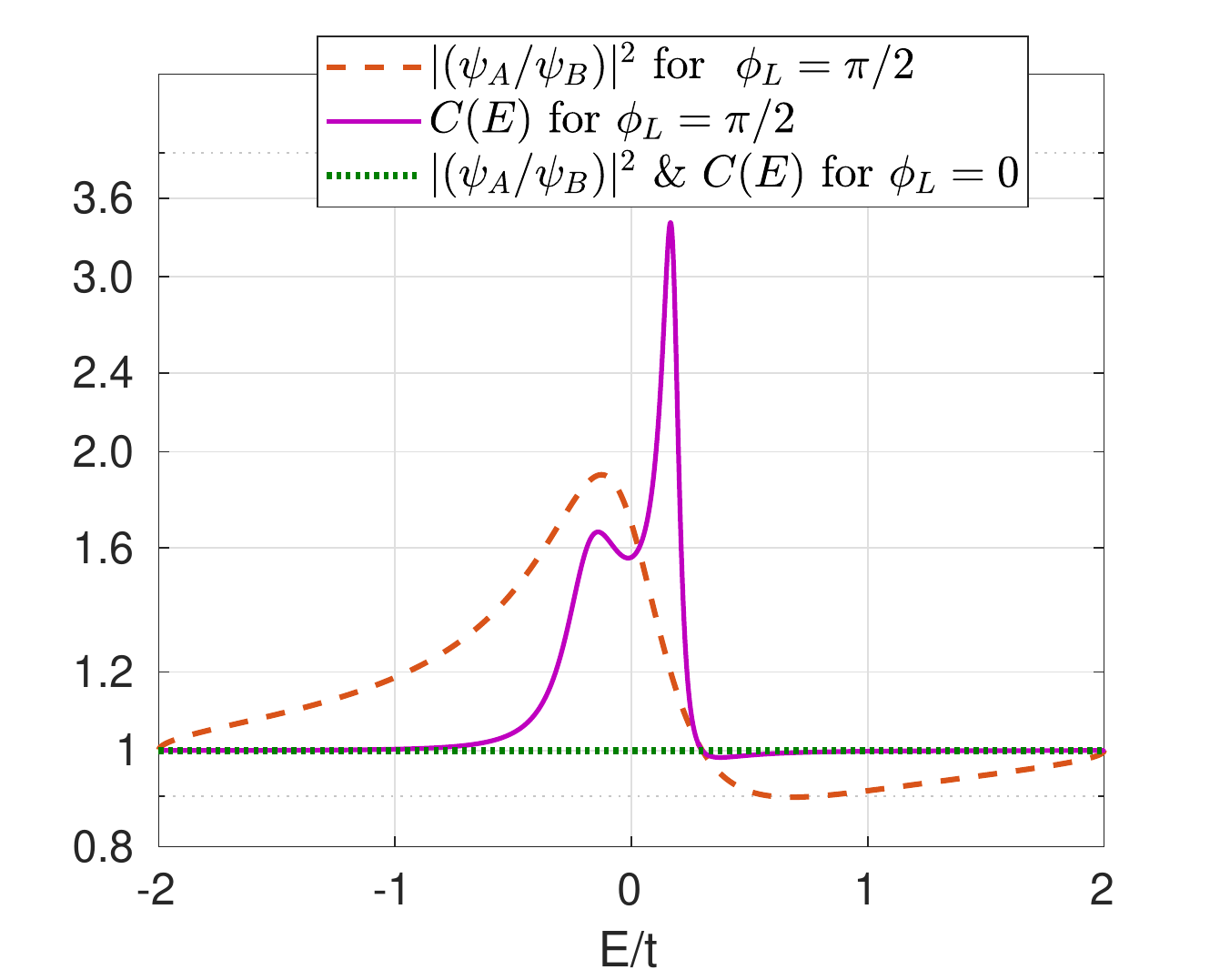}
 \caption{The sum of reflection and transmission probabilities $C(E)=|\mathbbm{r}_E|^2+|\mathbbm{t}_E|^2$ and $|(\psi_A/\psi_B)|^2$ plotted in log-scale as functions of energy $E$ for $t'=0.2t$, $\ga=0.1t$, $t_L=t_R=0.2t$, $\phi_R=0$. For $\phi_L\neq\phi_R$, current is not conserved  and $|\psi_A|\neq|\psi_B|$ at all energies. For $\phi_L=\phi_R$, the current is conserved and $|\psi_A|=|\psi_B|$ at all energies. }~\label{fig:cons}
\end{figure}
To begin with, let us analyze the current conservation for the scattering problem stated. For the model described by the Hamiltonian in eq.~\eqref{eq:ham}, a particle incident on the QD from $L$ has a wave function of the form eq.~\eqref{eq:psi}. For the choice of parameters: $t'=0.2t$, $\ga=0.1t$, $t_L=t_R=0.2t$, $\phi_L=0.5\pi$, $\phi_R=0$,  we plot $C(E)=|\mathbbm{t}_E|^2+|\mathbbm{r}_E|^2$ versus energy $E$ in Fig.~\ref{fig:cons}. We find that the probability current is not conserved at all energies. To get an insight into the reason behind this, we have also plotted the $|(\psi_A/\psi_B)|^2$ as a function of energy in Fig.~\ref{fig:cons} for the scattering wave function $\psi_n$. We find that whenever $|\psi_A|^2\neq|\psi_B|^2$, the probability current is not conserved and  $|\psi_A|^2\neq|\psi_B|^2$ whenever $\phi_L\neq\phi_R$. When $|\psi_A|^2>|\psi_B|^2$, the source is favored preferentially over the sink and $(|\mathbbm{r}_E|^2+|\mathbbm{t}_E|^2)>1$. On the other hand, when $|\psi_A|^2<|\psi_B|^2$, the sink is favored preferentially over the source and $(|\mathbbm{r}_E|^2+|\mathbbm{t}_E|^2)<1$, thus explaining the violation of probability current conservation. 
When $\phi_L=\phi_R$, we find that $|\mathbbm{r}_E|^2+|\mathbbm{t}_E|^2=1$ at all energies and the probability current is conserved. In this case,  we also find $|\psi_A|^2=|\psi_B|^2$ at all energies. 

\subsection{Transmission spectrum}
\begin{figure}[htb]
 \includegraphics[width=4.2cm]{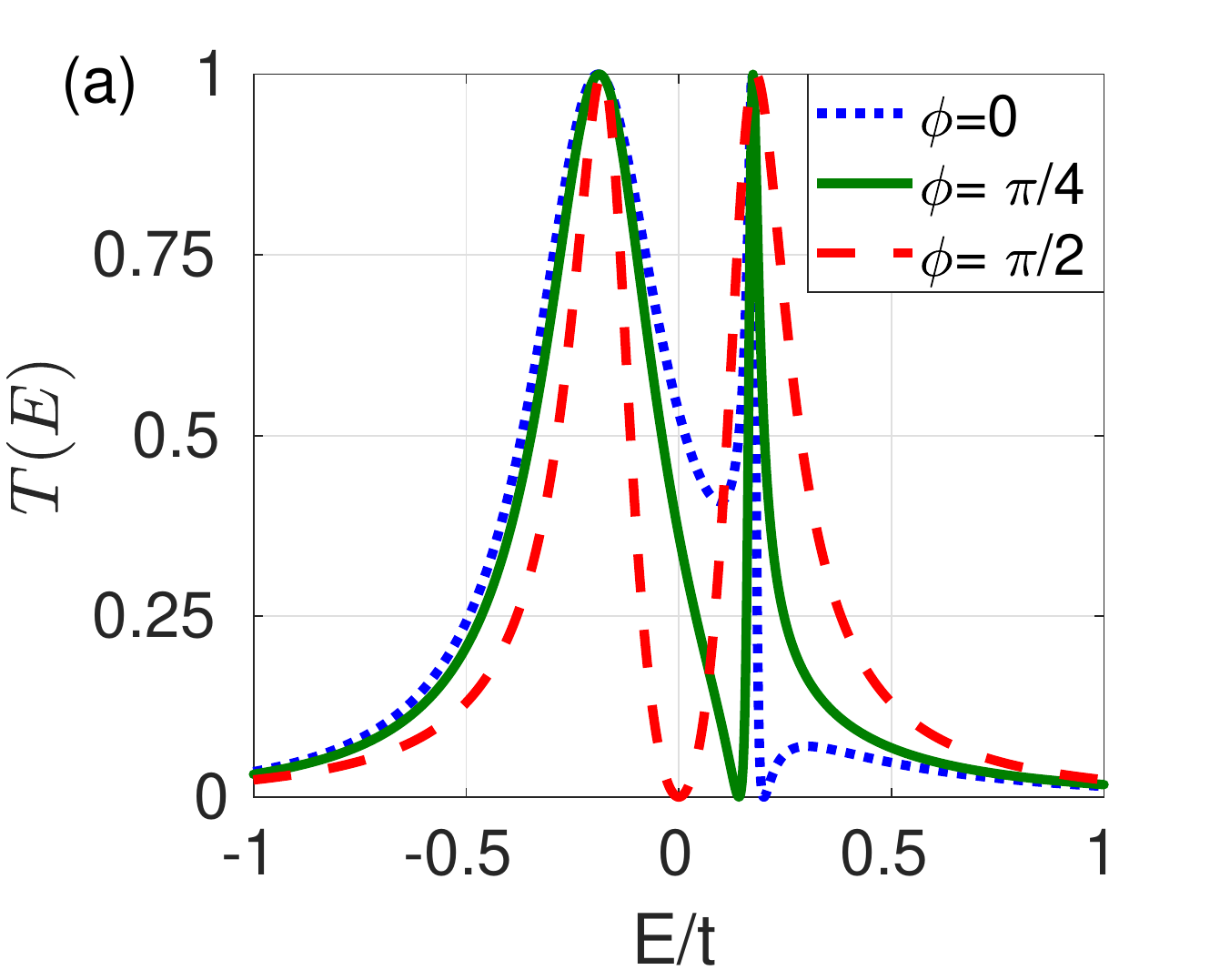}
 \includegraphics[width=4.2cm]{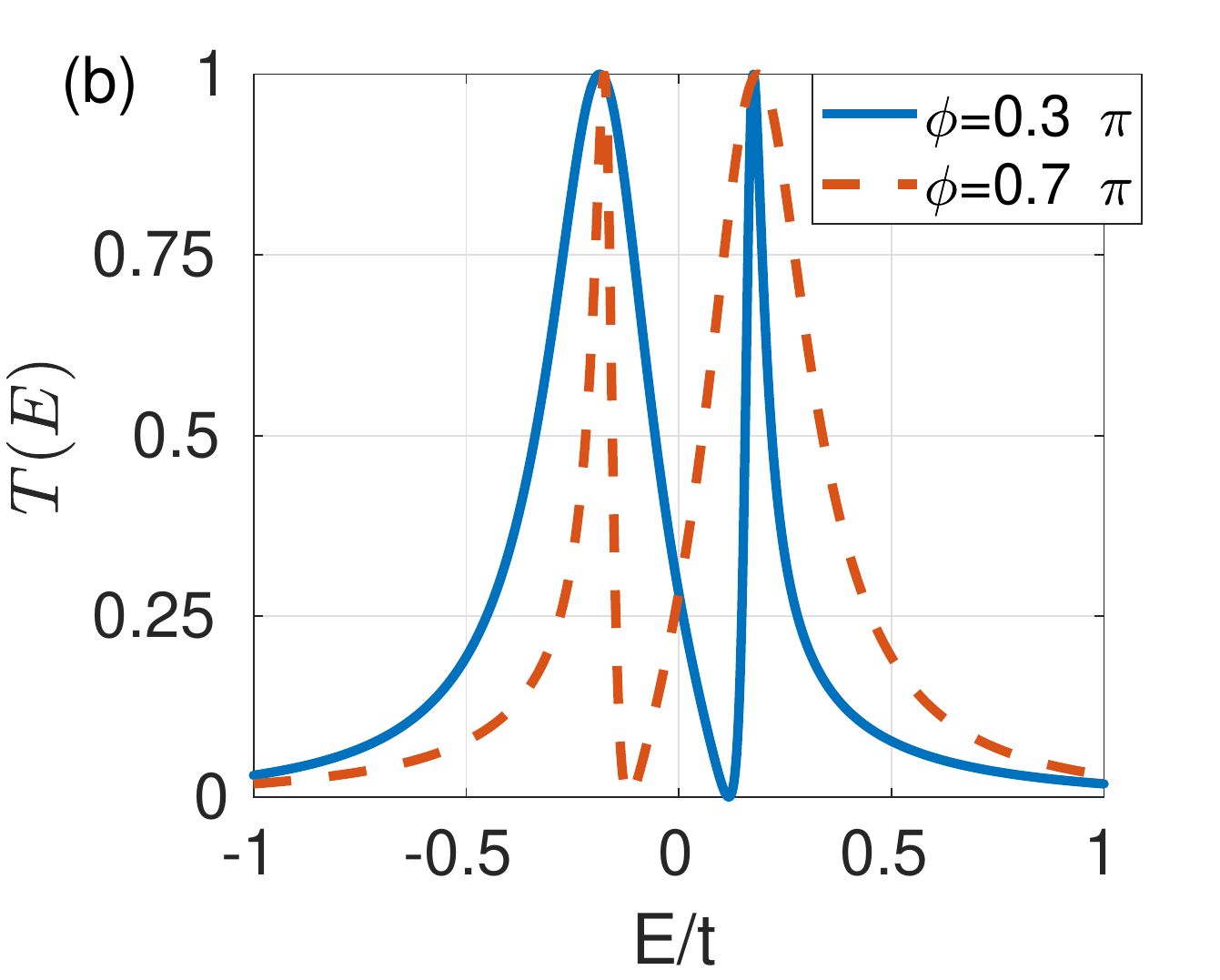}
 \caption{Transmission probability $T(E)=|\mathbbm{t}_E|^2$ versus energy $E$ for $t'=0.2t$, $\ga=0.1t$, $t_d=0.2t$. Value of $\phi$ for each curve is indicated in the legend. The values chosen for $\phi$ are different in (a) and (b), other parameters are the same. The plots in (a) show- `how the peak widths of the two peaks change with $\phi$?' The plots in (b) show that the peak widths of the two peaks get interchanged for the values of $\phi=\pi/2\pm\de \phi$. }~\label{fig:TvsE}
\end{figure}
Now, we study the dependence of the transmission probability on energy for different values of  the phase $\phi$, keeping other parameters the same. In Fig.~\ref{fig:TvsE}, we plot the transmission probability as a function of $E$ for various choices of $\phi$ mentioned in the legend. From the expression for $\mathbbm{t}_E$, it can be shown that $|\mathbbm{t}_E|^2=1$ at $E=E_{r,\pm}$ where
\bea E_{r,\pm}&=&[-t_d^2t'\cos{\phi}\pm\sqrt{\De}~]/(t^2-2t_d^2), ~ \nn \\ 
 \De &=& [t^2(t^2-2t_d^2)(t'^2-\ga^2)+t_d^4t'^2\cos^2{\phi}]. \label{eq:er}\eea For $t_d\ll t$, the condition $\De>0$ amounts to $\ga^2<t'^2$. This means that the eigenenergies of the isolated QD ($\ep_{\pm}$) are real, and the transmission probability is $1$ at these energies ($E_{r,\pm}\simeq\ep_{\pm}$). However, a glance at Fig.~\ref{fig:TvsE}~(a) indicates that the widths of the peaks at $E=E_{r,+}$ and $E=E_{r,-}$ are not the same for each of $\phi=0,\pi/4$. This can be explained by the fact that the overlap of the  eigenstate of the isolated QD $|\si\ra$ (whose eigenenergy is $\ep_{\si}$ for $\si=\pm$)  with the state $|\phi\ra=(|A\ra+e^{i\phi}|B\ra)/\sqrt{2}$ (which is coupled to the two semi-infinite lattices $L,~R$) is 
\bea |\la\phi|\si\ra|^2=t'(t'-\ga\sin{\phi}-\ep_{\si}\cos{\phi})/(2\ep^2_{\si}) \label{eq:ol} \eea 
Since the overlap is different for different eigenstates $|\si\ra$, the widths of the peaks are different. But the two overlaps are the same when $\phi=\pi/2$ implying that the widths of the two peaks should be the same for $\phi=\pi/2$, which agrees with the obtained result for $\phi=\pi/2$ as can be seen in Fig.~\ref{fig:TvsE}~(a). Also, the form of overlap suggests that the width of the peak at $E=E_{r,\si}$ for $\phi$ is the same as the width of the peak at $E=E_{r,-\si}$ when $\phi$ is replaced by $\pi-\phi$. We can see from Fig.~\ref{fig:TvsE}~(b) that this holds true. 

\begin{figure}[htb]
 \includegraphics[width=7.0cm]{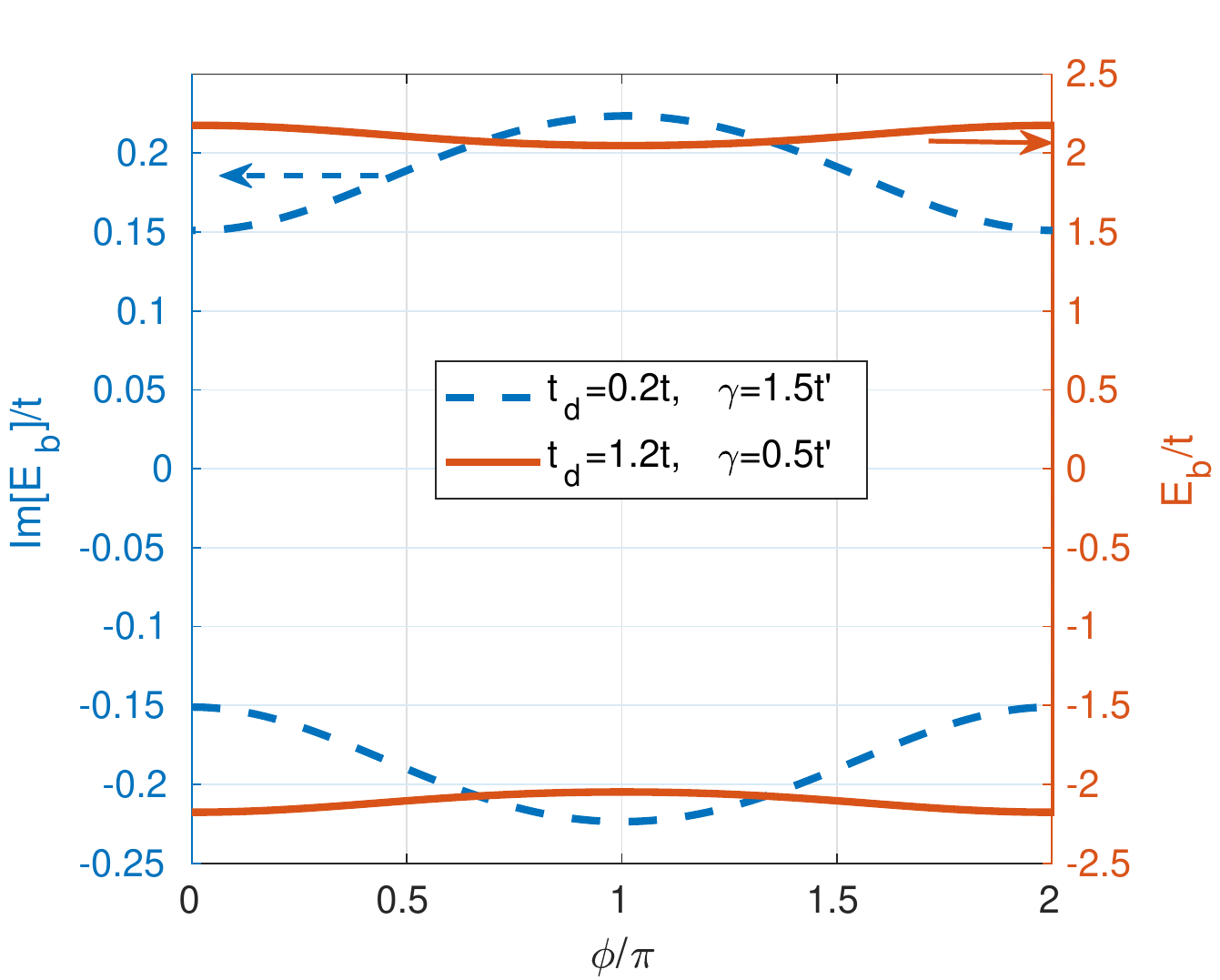}
 \caption{Bound state energies versus $\phi$. For $\ga=0.5t'$, $t'=0.2t$ and $t_d=1.2t$, bound state energies are real and are plotted as dot and dot-dash lines with values on the right ordinate. For $\ga=1.5t'$, $t'=0.2t$ and $t_d=0.2t$, the bound state energies are purely imaginary. The imaginary part of these energies are plotted on the left ordinate, represented by the solid and the dash lines. For all the curves, the abscissa is the same with $\phi$ as the corresponding variable. Other parameters are the same as in Fig.~\ref{fig:TvsE}. Since the scales for the real and imaginary energies are different, the real energy curves look flatter. }\label{fig:bound}
\end{figure}
A further scrutiny of eq.~\eqref{eq:er} gives us a complex value for $E_{r,\pm}$ for large $t_d$. This can be clearly seen in the limit $\phi=\pi/2$ and $t_d>t>0$. This means that the transmission probability is not exactly unity at even one energy in the range of energies of the semi-infinite lattices ($-2t<E<2t$). This motivates a search for bound states outside the energy range $-2t<E<2t$. A bound state wave function has the form 
\bea  \psi_n &=& \al_{s_n} e^{ik|n|},~~~{\rm for~~}|n|\ge 1, ~~\label{eq:psib} \eea
where $s_n={\rm sign}{(n)}$, $E=-2t\cos{k}$ and from the dispersion, $k$ is chosen so that ${\rm Im}[k]>0$ to make the wave function normalizable. For the bound state, the energy $E_b$ is either complex or $|E_b|>2t$ if real so that the imaginary part of $k$ is nonzero. By solving the Schr\"odinger equation numerically, we get two bound states.  For $t_d=1.2t$, $\ga=0.1t$ and $t'=0.2t$, we plot the bound state energies as functions of $\phi$ in Fig.~\ref{fig:bound}. We see that the bound state energies lie outside the range of energies of the two semi-infinite lattices. It is interesting to see that when $t_d>t$, even though the eigenenergies of the QD are real and lie within the band of the semi-infinite lattices, their coupling to the lattices makes them bound states with energy outside the range $-2t<E<2t$ and they do not participate in a scattering that results in perfect transmission. 

\subsection{$\mPT$-symmetry breaking of the quantum dot}
\begin{figure}[htb]
 \includegraphics[width=4.25cm]{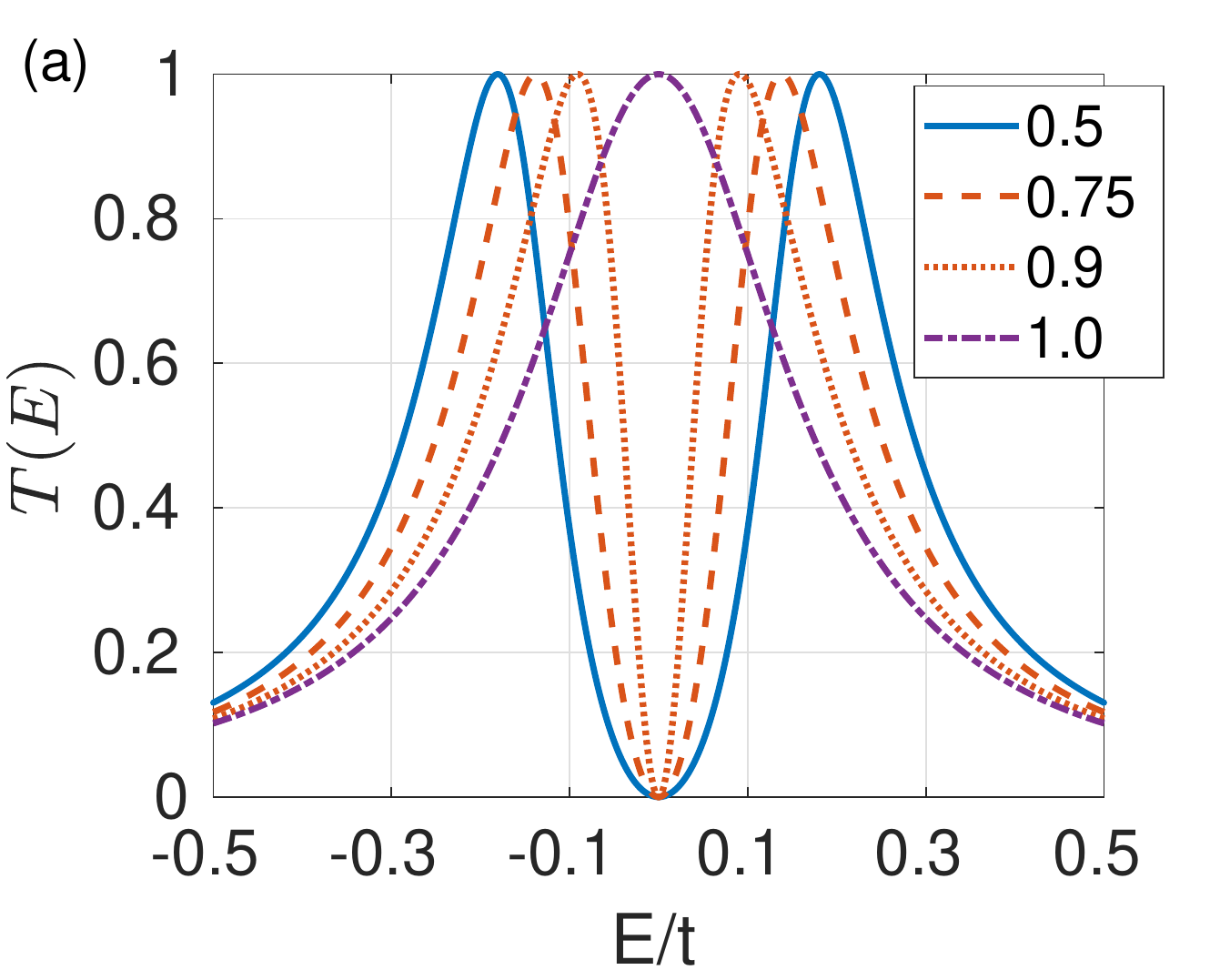}
 \includegraphics[width=4.25cm]{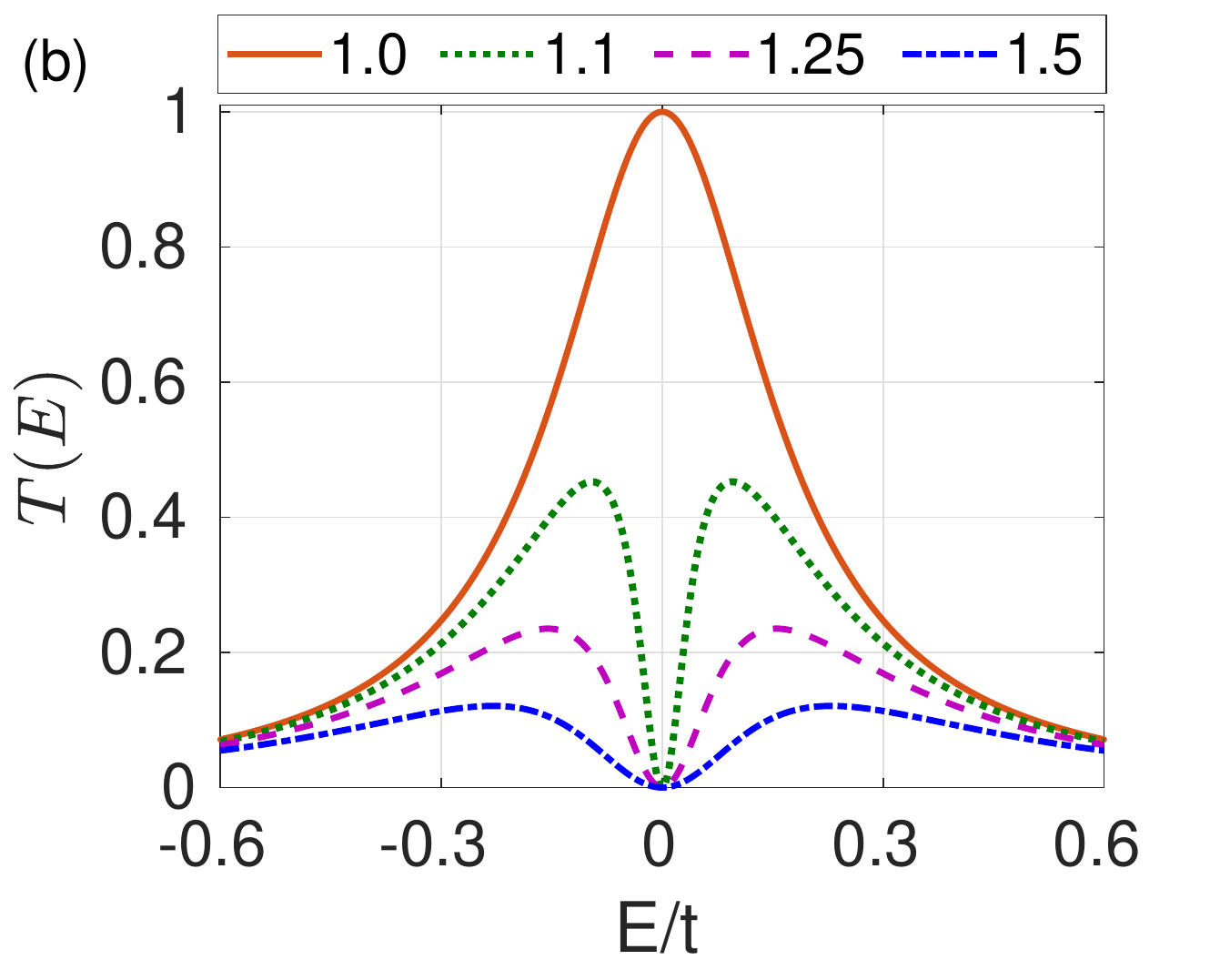}
 \caption{Transmission probability $T(E)=|\mathbbm{t}_E|^2$ versus energy for different values of $\ga/t'$ (indicated in the legend). (a) $\ga\le t'$ ($\mPT$-unbroken phase), (b) $\ga\ge t'$ ($\mPT$-broken phase). $\phi=\pi/2$ and the other parameters are the same as in Fig.~\ref{fig:TvsE}. In both the plots, the curve corresponding to $\ga=t'$ (exceptional point)  can be seen to produce one peak where the transmission is perfect.} ~\label{fig:TvsE-ga}
\end{figure}
Next, we choose $\phi=\pi/2$ and vary $\ga$ to see how the spectrum of transmission probability changes. For the isolated QD, the ratio $\ga/t'$ determines whether the eigenenergies are real ($\ga/t'\le 1$) or imaginary ($\ga/t'>1$). We plot the transmission probability versus energy in Fig.~\ref{fig:TvsE-ga} for different choices of $\ga/t'$ indicated in the legend. We see that as $\ga/t'\le 1$ approaches $1$, the positions of the two peaks get closer and at $\ga=t'$, the two peaks merge into one. For this range of $\ga$, the value of transmission probability at the peak is $1$. When $\ga/t'$ increases further from $1$, not only the positions of the two peaks move away from $E=0$, but also the value of transmission probability at the peak reduces. Absence of perfect transmission is because the eigenenergies of the isolated QD are purely imaginary, and they move further away from the real energy in the complex plane as $\ga/t'$ increases. Analytically, it can be shown that when $\phi=\pi/2$, for $\ga/t'>1$, the positions of the two peaks in the spectrum of transmission probability are at $E=E_{i,\pm}$ where 
\bea E_{i, \pm}=\pm\f{\sqrt{\ga^2-t'^2}}{\sqrt{1+(\ga^2-t'^2)/2t^2-2t_d^2/t^2}}, ~\label{eq:Ei}\eea  and the value of transmission probability at the peak is \bea |t_{E_{i, \pm}}|^2=\f{16t_d^4}{16t_d^4+(\ga^2-t'^2)^2+4(\ga^2-t'^2)(t^2-2t_d^2)}.~~\label{eq:TEi}\eea 

It can be seen from eq.~\eqref{eq:RT} that the transmission probability is proportional to $t_d^4$ for  $t_d\ll t$ for $\ga>t'$. Also, the energy $E_{i,\pm}$ at which the transmission probability is peaked does not change much with $t_d$ when $t_d\ll t$ [see eq.~\eqref{eq:Ei}]. The eigenenergies of the isolated QD ($\pm i\sqrt{\ga^2-t'^2}$) are complex, the energy of the incident electrons closest to these energies is $E=0$. Therefore, a first guess for the energy at which transmission probability would be maximum is $E=0$. This does not match the results obtained.

To get a further insight into why the transmission probability is peaked at the energies $E_{i,\pm}$, we look for bound states of the Hamiltonian in eq.~\eqref{eq:ham} choosing $\phi_L=\phi_R=\phi$ and $t_L=t_R=t_d$. 
We find that the real part of the bound state eigenenergy is zero for  $\ga>t'$ and small $t_d$. In Fig.~\ref{fig:bound}, we plot the imaginary part of the bound state energy as a function of $\phi$ fixing $\ga=1.5t'$ and keeping other parameters the same as before. This shows that the bound states with complex eigenenergies have real part equal to zero, and  nonzero imaginary part. This implies that at zero energy, the transmission probability should be peaked, since the energy of particles participating in scattering is real.  Hence, the bound state eigenenergies do not explain the peak in transmission probability at $E=E_{i,\pm}\neq 0$. To explain the peaks in transmission probability at $E=E_{i,\pm}\neq 0$, we design a four-site toy model in the next section. 

\subsection{Four-site toy model}

\begin{figure}[htb]
 \includegraphics[width=6cm]{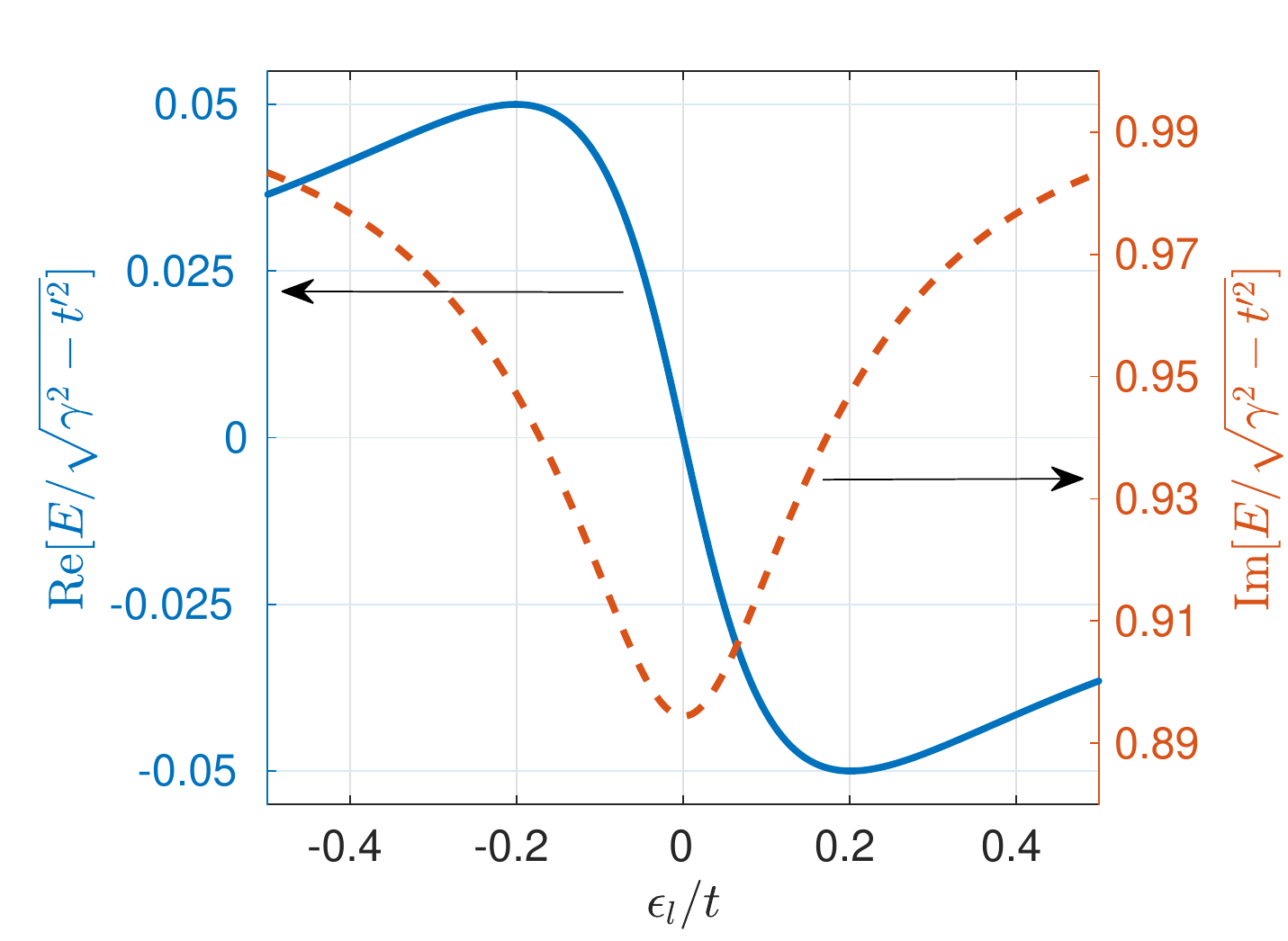}
 \caption{Real and imaginary parts of a complex eigenenergy $E$ of the four-site Hamiltonian $H_4$ as a function of $\ep_l$ for $t_d=0.05t$, $\phi=\pi/2$, $\ga=0.3t$ and $t'=0.2t$. The real part of the complex eigenenergy shows maximum deviation from zero at nonzero values of $\ep_l$. }\label{fig:el}
\end{figure}
Now, we study a toy model with four lattice sites closely related to the model described by eq.~\eqref{eq:ham}, in which two sites labelled $-1,~1$, with on-site energy $\ep_l$ are coupled to the QD. The Hamiltonian is given by: 
\bea
 H_4 &=& -t'(c^{\dag}_{A}c_B+h.c.)+i\ga (c^{\dag}_{A}c_A - c^{\dag}_{B}c_B) +\ep_l[c_{-1}^{\dag}c_{-1}\nn \\ && +c_1^{\dag}c_1]~-t_d[(c_{-1}^{\dag} + c_1^{\dag})(c_A+e^{i\phi}c_B) + h.c.]~~ \label{eq:h4}
 \eea
This is different from the Hamiltonian in eq.~\eqref{eq:ham} in the sense that in the latter case, two semi-infinite lattices are coupled to the QD while in the former case, it is just two sites that are coupled to the QD. From the construction of the Hamiltonian itself, one can see that one eigenstate of $H_4$ is $(|1\ra-|-1\ra)/\sqrt{2}$ with eigenenergy $\ep_l$. In the limit of small $t_d$, the eigenstates of the isolated QD hybridize with the state $(|1\ra+|-1\ra)/\sqrt{2}$
weakly. When $\ga>t'$, the eigenstates of the isolated QD come with purely imaginary eigenenergies, and they hybridize with the state $(|1\ra+|-1\ra)/\sqrt{2}$ to pick up a small real part which depends on $t_d$, $\ep_l$ and $\phi$.  We diagonalize the Hamiltonian $H_4$
numerically. There are at most two complex eigenvalues of $H_4$, which come in complex conjugate pairs.  We plot the real and imaginary parts of a complex eigenenergy as a function of $\ep_l$ in Fig.~\ref{fig:el} for the choice of parameters: $t_d=0.05t$, $\phi=\pi/2$, $\ga=0.3t$ and $t'=0.2t$.

\begin{figure}[htb]
 \includegraphics[width=4.25cm]{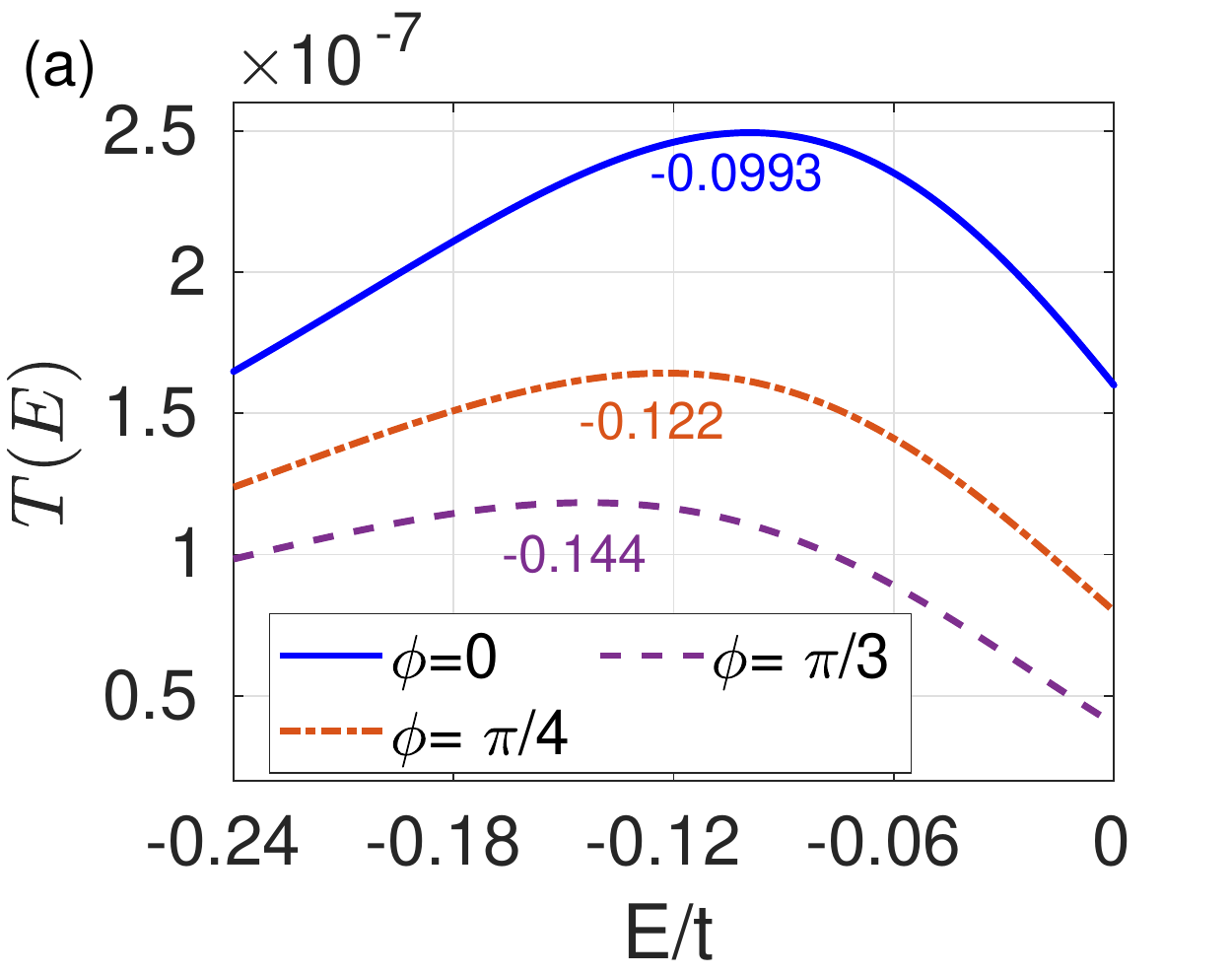}
 \includegraphics[width=4.25cm]{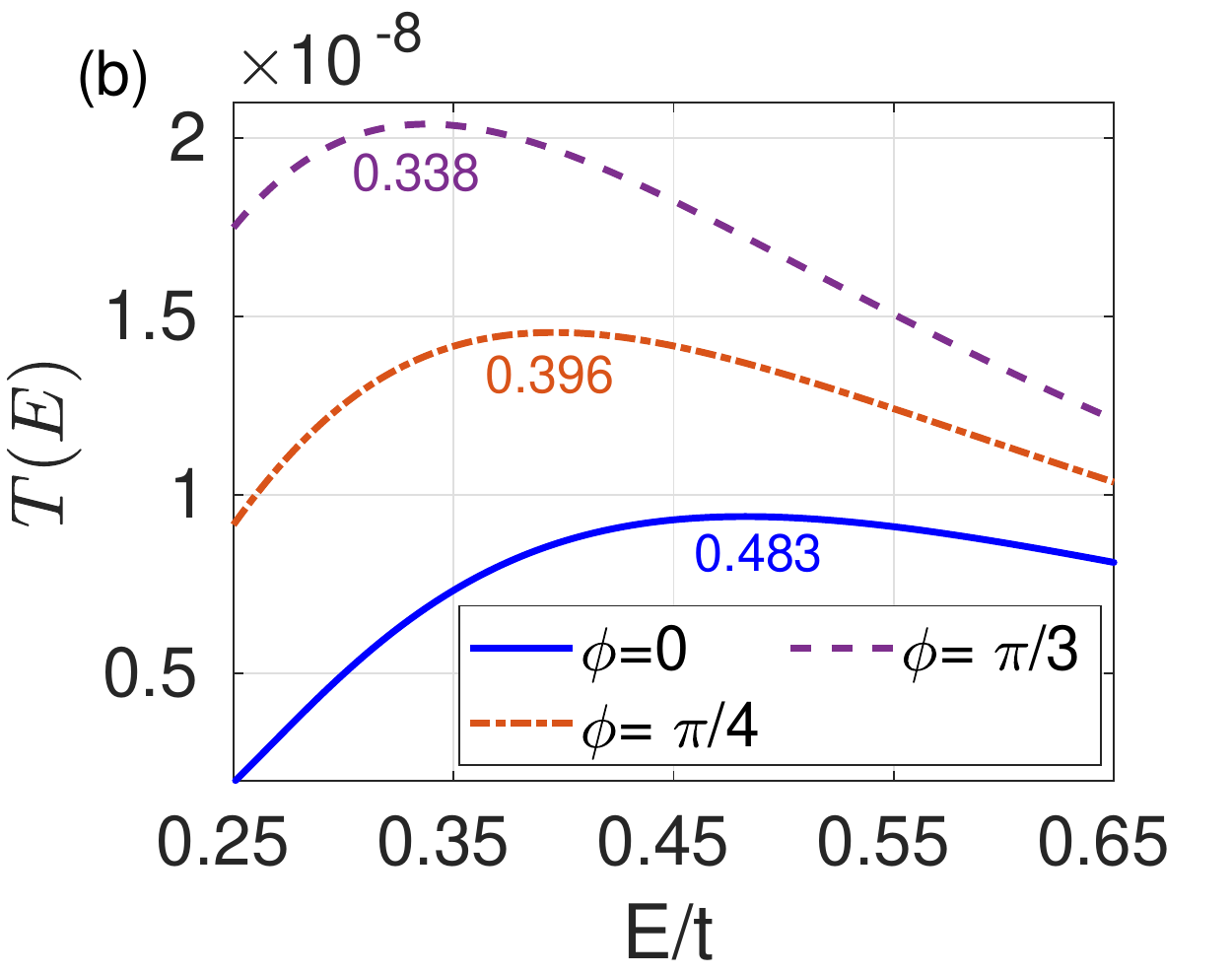}
 \includegraphics[width=4.25cm]{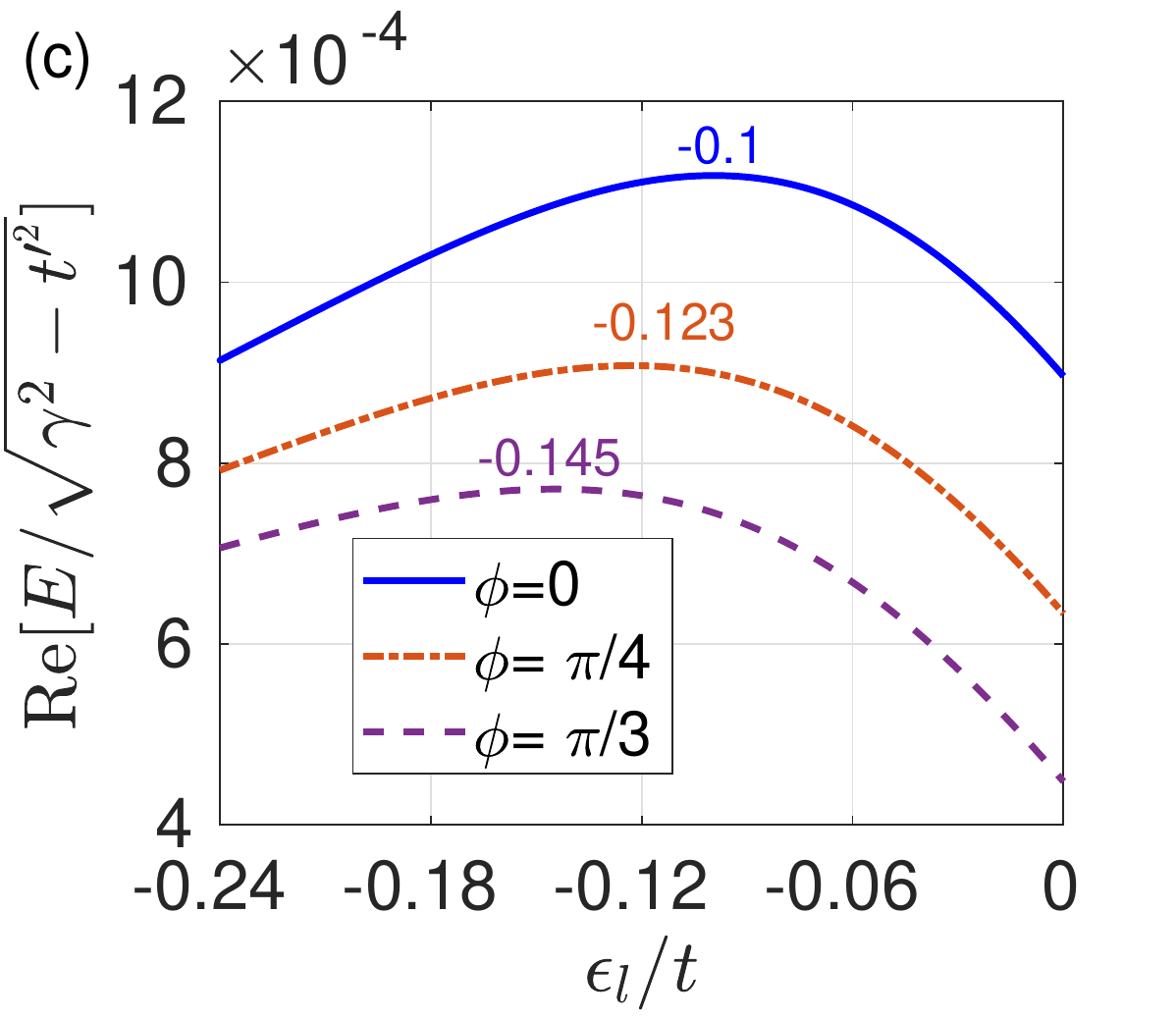}
 \includegraphics[width=4.25cm]{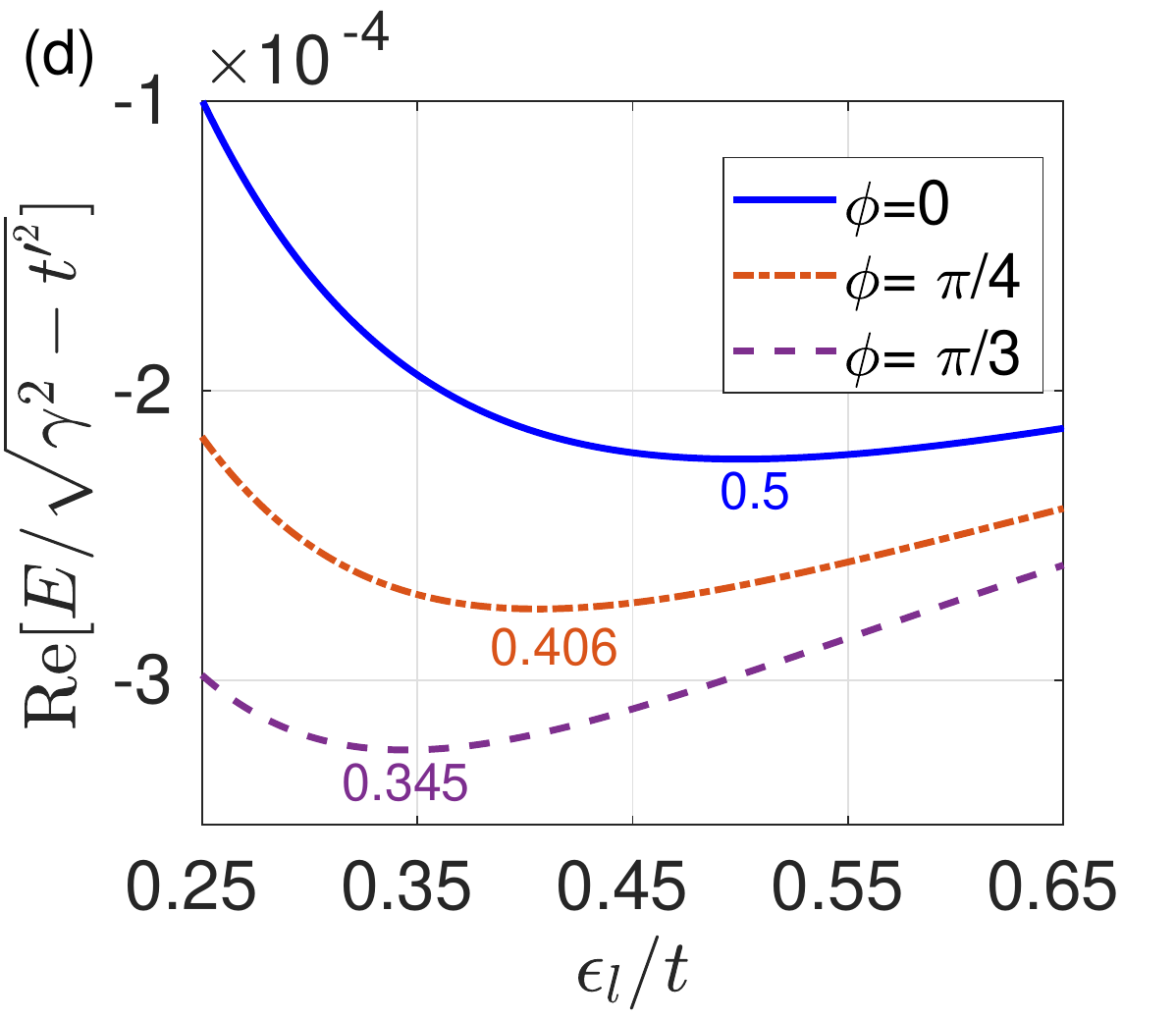}
 \caption{(a, b): Transmission probability $T(E)=|\mathbbm{t}_E|^2$ versus energy zoomed in near the two peaks. (c, d): Real part of a complex eigenenergy of $H_4$ as a function of $\ep_l$ zoomed in near the maximum deviation from zero. The value of $\phi$ corresponding to each plot is indicated in the legend. The numbers typed in the plots indicate the value of energy at which the function is shows a local extremum. Parameters: $t_d=0.005t$, $\ga=0.3t$ and $t'=0.2t$.  The values of $E$ in (a, b) at which $|\mathbbm{t}_E|^2$ is peaked matches well with the values of $\ep_l$ in (c, d) at which the deviation of the real part of a complex eigenenergy of $H_4$ is maximum.  }~\label{fig:TvsE-phi}
\end{figure}

We find that the shift of the real part of the complex  eigenenergies is maximum at some nonzero values of $\ep_l$. In this case, the maximum shift is around $\ep_l=\pm 0.2t$. If we take the limit as $t_d\to 0$, we get the maximum shift of the real part of the complex eigenenergies around $\ep_l= 0.2236$ which is the same as the value $\pm\sqrt{\ga^2-t'^2}$ ($\simeq E_{i,\pm}$ for  $\ga^2-t'^2\ll t^2$). This explains the peak in transmission probability at $E=E_{i,\pm}$ for $\ga>t'$.

Another feature of the transmission probability spectrum is that for $\phi=\pi/2$, we get perfect reflection at $E=0$ for $t'\neq \pm \ga$ (see Fig.~\ref{fig:TvsE-ga}). The system described by eq.~\eqref{eq:ham} can be thought of as two eigenstates of the isolated QD at energies $\pm\sqrt{t'^2-\ga^2}$ coupled to the two semi-infinite lattices with the same coupling [see eq.~\eqref{eq:ol}]. This is the same as the system described by  eq.~\eqref{eq:ham}, with the changes: $t'\to 0$ and $i\ga\to\sqrt{t'^2-\ga^2}$ (it can be seen from eq.~\eqref{eq:RT} that the scattering coefficients remain the same under these changes at $\phi=\pi/2$). It is known from eq.\eqref{eq:RT} that when $t'=0$ and $\ga \neq 0$, $\mathbbm{t}_E=0$ at $E=0$ making $|\mathbbm{r}_E|=1$. Perfect reflection at zero energy for $\phi=\pi/2$ is an effect of destructive interference between the paths through the two eigenstates of isolated QD.

Now, we turn to the case of $\phi\neq \pi/2$. In this case, the energy at which transmission probability is peaked changes from $E_{i,\pm}$. The plot of transmission probability versus energy is zoomed near the peak and presented in Fig.~\ref{fig:TvsE-phi}~(a, b) for the choice of parameters: $t_d=0.005t$, $\ga=0.3t$ and $t'=0.2t$. For the same set of parameters, the Hamiltonian $H_4$ [eq.~\eqref{eq:h4}] is diagonalised and the real part of the complex eigenvalues are plotted near the extrema in Fig.~\ref{fig:TvsE-phi}~(c, d). The values of the energy at which transmission probability is peaked and the values of the onsite energy $\ep_l$ at which the deviation of the real part of a complex eigenenergy is maximum (shown in the respective plots) agree well. 

\subsection{Aharonov-Bohm like interference}
\begin{figure}[htb]
 \includegraphics[width=8cm]{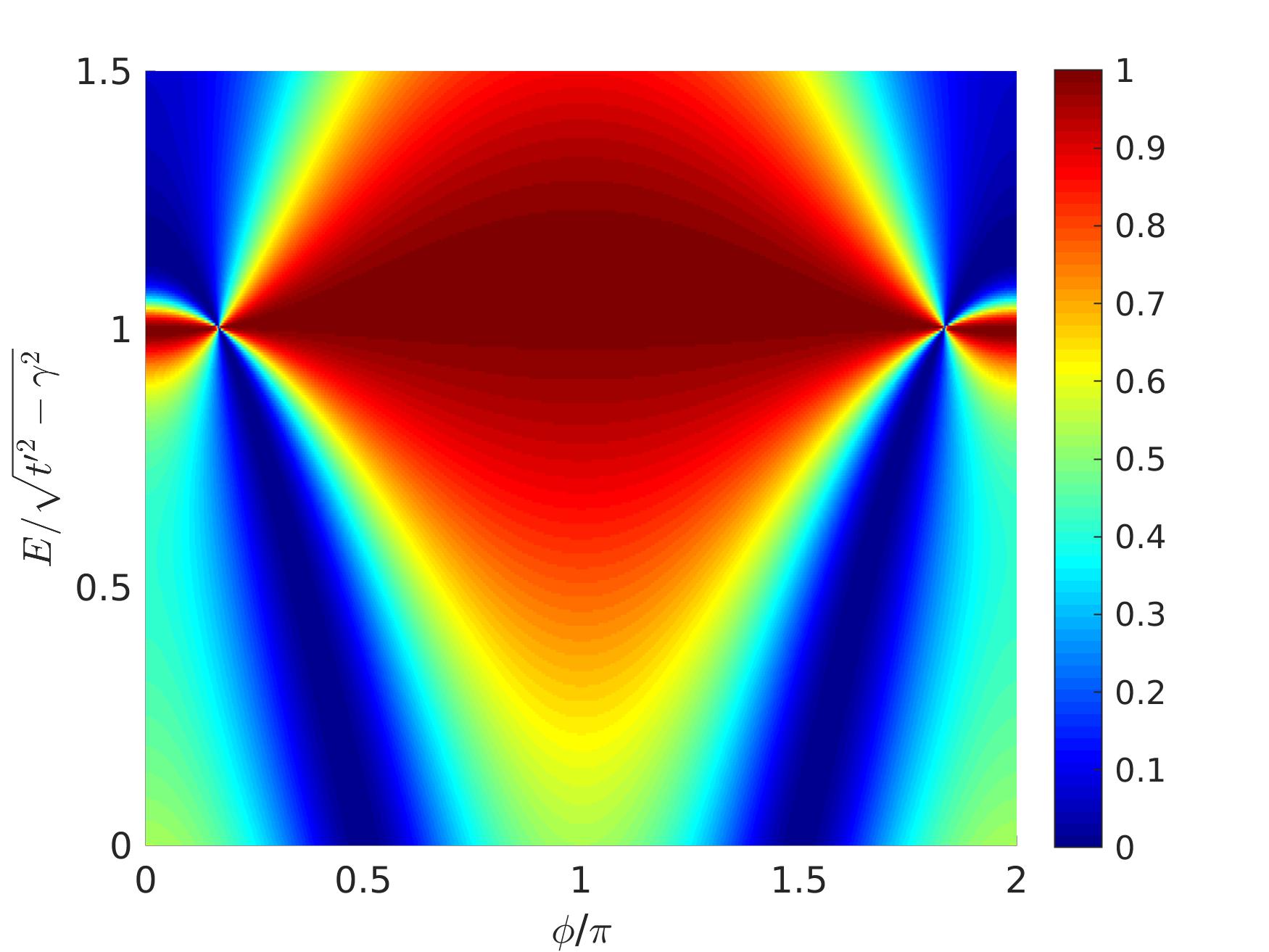}
 \caption{Aharonov-Bohm type interference: Transmission probability $T(E)=|\mathbbm{t}_E|^2$ as a function of $\phi$ and energy $E$.   Parameters: $t_d=0.2t$, $\ga=0.1t$ and $t'=0.2t$. For the choice of $E$ close to $\sqrt{t'^2-\ga^2}$, the transmission probability spans almost the entire range $[0,1]$ as $\phi$ is varied.} \label{fig:AB}
\end{figure}

Let us now examine the $\phi$-dependence of the transmission probability. In Fig.~\ref{fig:AB}, we plot the transmission probability versus $\phi$ and $E$  for the choice of parameters:  $t_d=0.2t$, $\ga=0.1t$ and $t'=0.2t$. We see that for energies $E$ close to $\sqrt{t'^2-\ga^2}$, the transmission probability is close to unity over a broad range of values of $\phi$. We can see from Fig.~\ref{fig:AB} that for $|E/\sqrt{t'^2-\ga^2}|$ around $1$, the transmission probability goes from $0$ to $1$ smoothly as $\phi$ changes. For other values of energy in the plot, the transmission probability does not span the entire range $[0,1]$. This is essentially an Aharonov-Bohm setup~\cite{aha59} with two quantum dots connected to reservoirs~\cite{akera93,fang10}, except that the flux due to $\phi$ is not in the entire closed loop between the two semi-infinite lattices. Here, a magnetic flux $\phi\hbar/e$ is enclosed in each of the two triangular loops: $(-1,A,B,-1)$ and $(1,A,B,1)$ (if the quantum particle has a charge $e$). The net flux in the loop $(-1,A,1,B,-1)$ is zero, which is essential for current conservation.
 
\section{Scattering across  $\mPT$-symmetric ladder}~\label{sec:ladder}

\subsection{Model and calculation}

We extend the idea of non-Hermitian $\mPT$-symmetric quantum dot to non-Hermitian $\mPT$-symmetric ladder. Connecting $N$ non-Hermitian $\mPT$-symmetric quantum dots makes a non-Hermitian $\mPT$-symmetric ladder of length $N$. The Hamiltonian for the ladder can be written as: 
\bea  
H_{0}&=& -t\sum_{n=1}^{N-1}(c^{\dag}_{0,n+1}c_{0,n}+h.c.) + \sum_{n=1}^{N}[-t'c^{\dag}_{0,n}\si_xc_{0,n}\nn \\&&+i\ga c^{\dag}_{0,n}\si_zc_{0,n} -\mu_0c^{\dag}_{0,n}c_{0,n}], ~~~~\label{eq:hladder}
\eea
where $c_{0,n}=[c_{A,n},~c_{B,n}]^T$. For the ladder, the operator $\mP$ takes the site $(A,n)$ to $(B,{N}-n+1)$ and the site $(B,n)$ to $(A,{N}-n+1)$ while the operator $\mT$ does complex conjugation making $\mPT H_0 \mPT = H_0$. 

\begin{figure}[htb]
 \includegraphics[width=8.5cm]{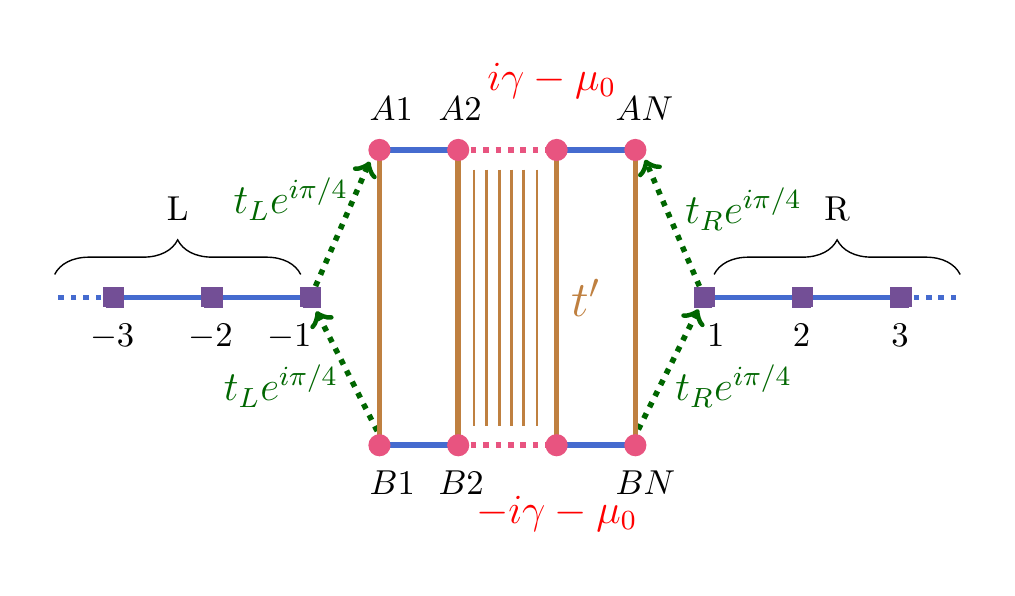}
 \caption{Schematic of a non Hermitian $\mPT$-symmetric ladder connected to semi-infinite lattices. }\label{fig:ladder}
\end{figure}

We connect such a ladder to semi-infinite lattices on two sides. The Hamiltonian for such a system can be written as: 
\bea 
 H ~~&=&~ H_L+H_{0}+H_R+H_{L0}+H_{0R}, ~~{\rm where} \nn \\
  H_{L0} &=& -t_L[c_{-1}^{\dag}(e^{-i\phi_L/2}c_{A,1}+e^{i\phi_L/2}c_{B,1}) + h.c.], \nn \\ 
   H_{0R} &=& -t_R[c_1^{\dag}(e^{-i\phi_R/2}c_{A, {N}}+e^{i\phi_R/2}c_{B, {N}})+h.c.],~~~~~ \label{eq:ham2}
\eea 
and $H_{L}$ and $H_R$ are the same as in eq.~\eqref{eq:ham}. The schematic diagram of the ladder connected to two semi-infinite lattices is shown in Fig.~\ref{fig:ladder}. The entire system is $\mPT$-symmetric only when $\phi_L=\phi_R$ and $t_L=r_R$ as will be discussed in sec~\ref{sec-con}.

The dispersion of the ladder is $E=-2t\cos{k}-\mu_0+\si\sqrt{t'^2-\ga^2}$, $\si=\pm$. This means that for  $\ga<t'$, the ladder is in $\mPT$-unbroken phase and for $\ga>t'$, the ladder is in $\mPT$-broken phase. Further, the limit $\ga=t'$ corresponds to the exceptional point. The scattering eigenstate  of a particle incident on the ladder from $L$ with energy $E$ ~($|\psi\ra=\sum_n\psi_n|n\ra$) has the same form as in  eq.~\eqref{eq:psi}, except that in addition to $|n|\ge 1$,  $n$ takes values $(A,1), (B,1), (A,2), (B,2),..., (A,{N}), (B,{N})$ for the sites on the ladder. For the sites on the ladder, $\psi_n$ can be written as $\psi_{0,m}=[\psi_{A,m},~\psi_{B,m}]^T$ which has the form   
 \bea
 \psi_{0, m} &=& \sum_{\si, \nu=\pm} s_{E,\si,\nu} e^{i\nu k_{\si}m}[u_{\si}, ~v_{\si}]^T,  ~{\rm for ~}1\le m\le {N},~~~~~~~  \label{eq:psiLA} 
 \eea
where $u_{\si}=t'$ and $v_{\si}=i\ga-\si \sqrt{t'^2-\ga^2}$. The scattering coefficients $\mathbbm{r}_E, \mathbbm{t}_E, s_{E,\si,\nu}$ can be determined using Schr\"odinger wave equation.  

\subsection{Results for transmission across the ladder}

\begin{figure}[htb]
 \includegraphics[width=9cm]{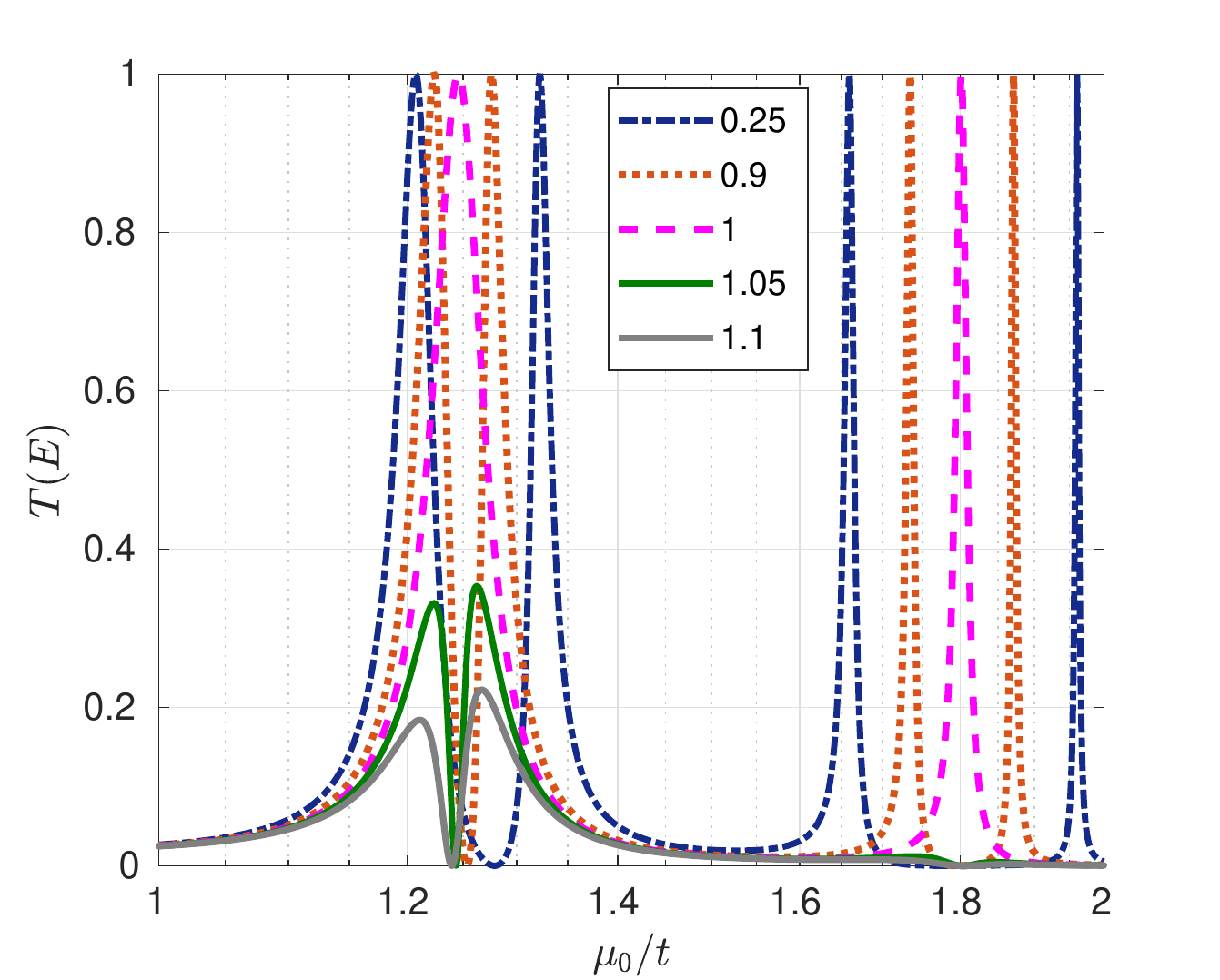}
 \caption{Transmission across  ladder: Transmission probability $T(E)=|\mathbbm{t}_E|^2$ at zero energy versus the chemical potential $\mu_0$ for $N=6$, $t_L=t_R=0.2t$ and $t'=0.2t$ for different values of $\ga/t'$ indicated in the legend. Abscissa is plotted in log-scale for better clarity. In the $\mPT$-unbroken phase $|\mathbbm{t}_E|^2=1$ for some values of $\mu_0$ whereas in the $\mPT$-broken phase, $|\mathbbm{t}_E|^2<1$ for all values of $\mu_0$. }~\label{fig:Tladder}
\end{figure} 

It can be analytically shown  that the probability current is conserved for scattering across the ladder only when $\phi_L=\phi_R$. We choose $\phi_L=\phi_R=\pi/2$, solve for the scattering amplitudes numerically for $N=6$, $t'=0.2t$, $t_L=t_R=t_d=0.2t$ and plot transmission probability at zero energy as a function of $\mu_0$ in Fig.~\ref{fig:Tladder}. For $\ga<t'$, the isolated ladder region has real energies while for $\ga>t'$, the eigenenergies of the isolated ladder region are complex with nonzero imaginary part. We can see that for $\ga\le t'$, the transmission probability can be tuned to unity  at specific values of $\mu_0$. On the other hand, for $\ga>t'$, the transmission probability is never unity for any value of $\mu_0$. This is because the dispersion relation for the ladder gives a complex spectrum when $\ga>t'$ and for a particle incident from a lattice with real energy, the corresponding wave numbers $k_{\si}$ are complex, making the wave function decay into the ladder away from the junction with the semi-infinite lattice. For $\ga\le t'$, the perfect transmission at different values of $\mu_0$ is due to Fabry-P\'erot interference~\cite{soori12,soori17,nehra19,soori19,suri21,suri21,soori21phesoc,soori22car} of the plane wave modes in the ladder region. It can also be seen that the number of peaks at the exceptional point ($\ga=t'$) is half the number of peaks away from the exceptional point, and the value of transmission probability at the peak is $1$ at the exceptional point.  Further, for fixed values of $\mu_0$ and $E$, the transmission probability  oscillates with $N$ for $\ga\le t'$ whereas it decays exponentially with $N$ for $\ga>t'$.

\section{Discussion and Conclusion}~\label{sec-con}
In this work, we have studied scattering in systems where semi-infinite one dimensional lattices are connected to $\mPT$-symmetric quantum dot and $\mPT$-symmetric ladder in such a way that the probability current is conserved. While the operator $\mP$ interchanges the  sites $A$ and $B$, when applied to sites of the one-dimensional semi-infinite lattice,  takes the site $n$ to $-n$ (for $|n|\ge 1$). This makes the entire system $\mPT$-symmetric only when $t_L=t_R$ and $\phi_L=\phi_R$. While $\phi_L=\phi_R$ is necessary for the probability current to be conserved, $t_L$ need not be equal to $t_R$ for current conservation. Also, when $\phi_L=\phi_R$,  the transmission and the reflection probabilities are the same for particles incident on the QD from the left and from the right,  akin to scattering of noninteracting particles across a Hermitian quantum dot~\cite{roy09}. The table~\ref{tab1} below summarizes these results for different cases of the models in eq.~\eqref{eq:ham} and eq.~\eqref{eq:ham2} studied here.  It can be seen that while $\mPT$-symmetry guarantees conservation of probability current, the conservation of probability current does not imply that the system is $\mPT$-symmetric. 
\begin{table}[htb]
 \begin{center}
\begin{tabular}{| c| c |c |}
\hline
Parameters  & $\mPT$-symmetry & Probability current  \\ 
\hline
 $\phi_L\neq\phi_R$ & No & Not conserved \\  
 \hline
  $\phi_L=\phi_R$, $t_L=t_R$& Yes & Conserved \\
 \hline
  $\phi_L=\phi_R$, $t_L\neq t_R$& No & Conserved \\
 \hline
\end{tabular}
\end{center}
\caption{Different cases of the model described by eq.~\eqref{eq:ham} where the $\mPT$-symmetry of the entire system and the conservation of probability current are tabulated.}\label{tab1}
\end{table}
Whenever $\phi_L=\phi_R$, not only the current is conserved, the S-matrix is unitary and the phenomenon of unidirectional invisibility does not show up.

 Recently, in pseudo-Hermitian systems, the current is found to be conserved~\cite{jin22}. A Hamiltonian $H$ is called pseudo-Hermitian if  $H^{\dag}=U^{\dag}HU$ where $U$ is a unitary operator. The previous lattice models~\cite{jin12,zhu15} wherein the current is conserved are pseudo-Hermitian systems.  The Hamiltonians in eq.~\eqref{eq:ham} and eq.~\eqref{eq:ham2} we proposed are  pseudo-Hermitian only in the limit $\phi_L=\phi_R=0$. Hence, our work points to a more general condition for current conservation.

In summary, we have studied a model where the probability current is conserved in  scattering across a non-Hermitian $\mPT$-symmetric region. For a $\mPT$-symmetric quantum dot weakly connected to two lattices, the transmission probability is unity at two energies close to eigenenergies of the isolated QD in the $\mPT$-unbroken phase. We have obtained analytical expression for energies at which the transmission is perfect.  In the $\mPT$-broken phase, the transmission probability is always strictly less than $1$. However, the transmission probability is peaked at two energies. A four-site toy model is used to quantitatively explain this feature. At the exceptional point, the transmission probability is $1$ at only one value of energy which is equal to the eigenenergy of the QD at the exceptional point.  We have discussed the bound states for the system in the $\mPT$-broken and $\mPT$-unbroken phases. Further, the parameter $\phi$, which physically corresponds to  magnetic fluxes threading through triangular loops, influences the transmission probability. This is an interference phenomenon closely related to Aharonov-Bohm effect~\cite{aha59,akera93}. In our setup, the two adjacent triangular loops  formed by the quantum dot with the lattice sites on the left and the right (in clockwise direction) enclose magnetic flux $\phi\hbar/e$ and $-\phi\hbar/e$ respectively. In an experimental setup where such a QD is connected to two semi-infinite lattices,  piercing fluxes $\phi\hbar/e$ and $-\phi\hbar/e$ through neighboring loops can be challenging. However, placing two one dimensional lattices on $x$-axis ($L$ in $x<0$ and $R$ in $x>0$), with the site $A$ at $\vec r =(0,1,1)$ and the site $B$ at $\vec r = (0,-1,1)$, and applying a magnetic field along $\hat x$ can mimic the setup described by eq.~\eqref{eq:ham}. In a variety of platforms  $\mathcal{PT}$-symmetric non-Hermitian systems have been experimentally realized~\cite{guo09,ruter,kottos10,longhi17,bittner,sun14,poli15,schindler,zhu14}. A tight binding lattice with non Hermitian onsite potentials has been engineered experimentally~\cite{poli15}, and the model proposed here in our work  can be possibly realized. Optical  waveguide structures with balanced loss and gain have been realized~\cite{Ctyroky10,ruter} and such platforms provide an avenue where the theoretical ideas discussed in our work can be put to test. 

%

\acknowledgements
We thank Diptiman Sen for comments on the manuscript and Adhip Agarwala for useful discussions. 
 A.~S. acknowledges DST, India for financial support through DST-INSPIRE Faculty Award (Faculty Reg. No.~:~IFA17-PH190). V.~S. acknowledges SERB (SERB/F/5343/2020-21) for financial support through MATRICS scheme. 
\bibliography{refnh}

\end{document}